\documentclass[journal,12pt,onecolumn,draftclsnofoot,]{IEEEtran}

\usepackage[cmex10]{amsmath}
\usepackage[utf8]{inputenc}
\usepackage{multirow}
\usepackage{soul}
\usepackage{algorithm}
\usepackage{algpseudocode}
\usepackage{graphicx}
\usepackage{dsfont}
\usepackage{xcolor}
\usepackage{cite}

\usepackage[utf8]{inputenc}
\usepackage{mathtools}
\usepackage[mathscr]{euscript}
\usepackage{amsmath}
\usepackage{amssymb}
\usepackage{amsfonts}
\usepackage{amssymb}
\usepackage{amsmath}
\usepackage{verbatim}
\usepackage[T1]{fontenc}
\usepackage[utf8]{inputenc}
\usepackage{latexsym}
\usepackage{enumerate}
\usepackage{indentfirst}
\usepackage{calligra}
\usepackage{ulem}
\usepackage{graphicx}
\usepackage{array}
\usepackage{float}
\usepackage{bbm}

\usepackage{geometry}

\usepackage{mleftright}
\usepackage{colortbl}
\makeatletter
\let\old@ps@headings\ps@headings
\let\old@ps@IEEEtitlepagestyle\ps@IEEEtitlepagestyle
\def\psccfooter#1{%
 \def\ps@headings{%
 \old@ps@headings%
 \def\@oddfoot{\strut\hfill#1\hfill\strut}%
 \def\@evenfoot{\strut\hfill#1\hfill\strut}%
 }%
 \def\ps@IEEEtitlepagestyle{%
 \old@ps@IEEEtitlepagestyle%
 \def\@oddfoot{\strut\hfill#1\hfill\strut}%
 \def\@evenfoot{\strut\hfill#1\hfill\strut}%
 }%
 \ps@headings%
}
\makeatother

\usepackage{soul}
\usepackage{multirow}
\usepackage{soul,color}
\usepackage{graphicx}
\usepackage{dsfont}
\usepackage{subcaption}

\usepackage[utf8]{inputenc}
\usepackage{mathtools}
\usepackage[mathscr]{euscript}
\usepackage{amsmath}
\usepackage{amssymb}
\usepackage{amsfonts}
\usepackage{verbatim}
\usepackage[T1]{fontenc}
\usepackage[utf8]{inputenc}

\usepackage{latexsym}
\usepackage{enumerate}
\usepackage{indentfirst}
\usepackage{calligra}
\usepackage{ulem}
\usepackage{multirow}

\usepackage{array}
\usepackage{float}
\usepackage{bbm}

\usepackage{eurosym}

\usepackage{hyperref}

\usepackage{caption}
\usepackage{tikz}
\usepackage{pgfplots}

\pgfplotsset{compat=1.8}
\usepgfplotslibrary{statistics}
\pgfmathdeclarefunction{fpumod}{2}{%
 \pgfmathfloatdivide{#1}{#2}%
 \pgfmathfloatint{\pgfmathresult}%
 \pgfmathfloatmultiply{\pgfmathresult}{#2}%
 \pgfmathfloatsubtract{#1}{\pgfmathresult}%
 \pgfmathfloatifapproxequalrel{\pgfmathresult}{#2}{\def\pgfmathresult{5}}{}%
 }

\usepackage{tabularx}
\usepackage{flushend}
\usepackage{textcomp}
\usepackage{xcolor}
\usepackage{import}

\usetikzlibrary{trees,decorations,shadows}
\tikzset{level 1/.style={sibling angle=45,level distance=4mm}}
\usetikzlibrary{arrows.meta}
\usepackage{forest}
\usetikzlibrary{external}

\let\oldtikzexternalgetnextfilename\tikzexternalgetnextfilename \renewcommand{\tikzexternalgetnextfilename}[1]{\oldtikzexternalgetnextfilename{#1}\expandafter\tikzsetnextfilename\expandafter{#1}}

\usepackage{pgfplotstable}
\usepackage[outline]{contour}
\contourlength{1.2pt}

\pgfplotsset{compat=1.13} 

\usetikzlibrary{spy}
\usetikzlibrary{calc}
\usetikzlibrary{fadings}
\usetikzlibrary{patterns}
\usetikzlibrary{shadows}
\usetikzlibrary{mindmap}
\usetikzlibrary{backgrounds}
\usetikzlibrary{shapes.symbols}
\usetikzlibrary{shapes.multipart}
\usetikzlibrary{shapes.geometric}
\usetikzlibrary{automata,positioning}
\usetikzlibrary{decorations.fractals} 
\usetikzlibrary{decorations.markings}
\usetikzlibrary{decorations.pathreplacing}
\usetikzlibrary{decorations.pathmorphing}
 
\usepackage{bm}

\usepackage{nomencl}
\makenomenclature
\tikzset{edge from parent/.style={segment angle=10,draw}}

\tikzset{
 my rounded corners/.append style={rounded corners=2pt},
}

\def\BibTeX{{\rm B\kern-.05em{\sc i\kern-.025em b}\kern-.08em
 T\kern-.1667em\lower.7ex\hbox{E}\kern-.125emX}}

\renewcommand{\nomgroup}[1]{%
 \ifthenelse{\equal{#1}{O}}{\item[\textit{Operators}]}{%
 \ifthenelse{\equal{#1}{I}}{\item[\textit{Indices}]}{%
 \ifthenelse{\equal{#1}{A}}{\item[\textit{Acronyms}]}{%
 `\ifthenelse{\equal{#1}{V}}{\item[\textit{Variables and parameters}]}{}}}}}
\usepackage{scalerel}
\usetikzlibrary{svg.path}
\definecolor{orcidlogocol}{HTML}{A6CE39}
\tikzset{
 orcidlogo/.pic={
 \fill[orcidlogocol] svg{M256,128c0,70.7-57.3,128-128,128C57.3,256,0,198.7,0,128C0,57.3,57.3,0,128,0C198.7,0,256,57.3,256,128z};
 \fill[white] svg{M86.3,186.2H70.9V79.1h15.4v48.4V186.2z}
 svg{M108.9,79.1h41.6c39.6,0,57,28.3,57,53.6c0,27.5-21.5,53.6-56.8,53.6h-41.8V79.1z M124.3,172.4h24.5c34.9,0,42.9-26.5,42.9-39.7c0-21.5-13.7-39.7-43.7-39.7h-23.7V172.4z}
 svg{M88.7,56.8c0,5.5-4.5,10.1-10.1,10.1c-5.6,0-10.1-4.6-10.1-10.1c0-5.6,4.5-10.1,10.1-10.1C84.2,46.7,88.7,51.3,88.7,56.8z};
 }
}

\newcommand\orcidicon[1]{\href{https://orcid.org/#1}{\mbox{\scalerel*{ \begin{tikzpicture}[yscale=-1,transform shape]
 \pic{orcidlogo};
 \end{tikzpicture}
 }{|}}}}

\begin{document}
%
\title{{Robust dynamic operating envelopes for flexibility operation {using only local voltage measurement}}}

\author{Md~Umar~Hashmi$^{1}$*\orcidicon{0000-0002-0193-6703},
and~Dirk~Van~Hertem$^{1}$~\orcidicon{0000-0001-5461-8891}
\thanks{Corresponding author email: mdumar.hashmi@kuleuven.be}
\thanks{$^{1}$Md Umar Hashmi and Dirk Van Hertem are with KU Leuven, division Electa \& EnergyVille, Genk, Belgium}
\thanks{This work is supported by 
the H2020 EUniversal project, grant agreement ID: 864334 (\url{https://euniversal.eu/}) and 
the Flemish Government and Flanders Innovation \& Entrepreneurship (VLAIO) through the Moonshot project InduFlexControl (HBC.2019.0113) and project IMPROcap (HBC.2022.0733).
}}

\maketitle


\begin{abstract}
With growing intermittency and uncertainty in distribution networks around the world, ensuring operational integrity is becoming challenging.
Recent use cases of dynamic operating envelopes (DOEs) indicate that \textcolor{black}{they} can be utilized for network awareness for autonomous operation of flexibility, maximizing distributed generation integration, coordinating flexibility in different power networks and in resource planning. 
\textcolor{black}{To this end, a novel framework is presented for generating decentralized DOEs in real-time using only the nodal voltage measurement and partially decentralized, risk-averse, robust DOEs  in a time-ahead setting using voltage forecast scenarios.}
Chance constraint level is analytically implemented for avoiding extremely restrictive time-ahead DOEs with insufficient feasible regions for local energy optimization.
Since the proposed DOE calculation framework uses \textcolor{black}{none or limited} centralized feedback, it is resilient to cyberattacks, communication failures, missing data and errors in network layout information.
\textcolor{black}{
Numerical results showcase the DOE calculation framework in real-time using voltage magnitude measurements and in day-ahead timeframe using forecasted voltage scenarios.} Furthermore, the DOEs are extended to form P-Q charts while considering power factor and converter capacity limits.
\end{abstract}


\pagebreak

\tableofcontents

\pagebreak


\section{Introduction}
\begin{figure}[!htbp]
	\centering
	\includegraphics[width=0.9\columnwidth]{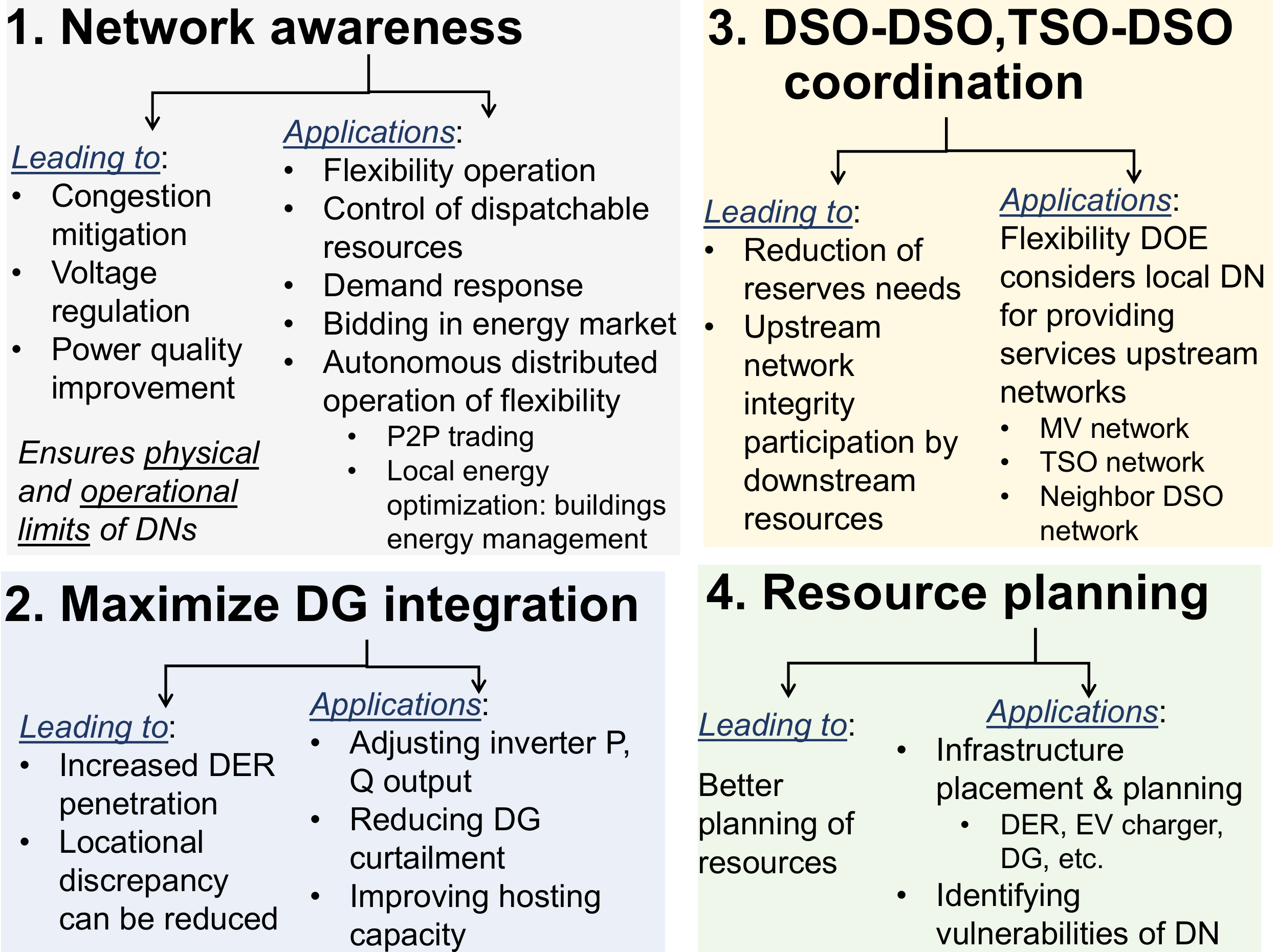}
	\vspace{-4pt}
	\caption{{Use cases for DOEs in distribution and transmission networks.} }
	\label{fig:usecase}
\end{figure}

Flexible resources participating in energy markets would  impact the operating states of the distribution network (DN) to which it is connected to. Dynamic operating envelopes (DOE) provide a feasible range for operating flexible resources while considering local DN objectives for preserving the network integrity by ensuring voltage and line loading are within permissible bounds.
The use cases for DOE are detailed in Fig. \ref{fig:usecase}.
Based on prior literature, these use cases are classified into four objectives: (1) providing network operation for autonomous operation, (2) maximizing distributed generation (DG) integration to the DNs, (3) applicability of DN flexibility resources for solving upstream network issues in medium and high voltage networks and also for neighbouring DNs, and (4) use case of DOE for resource planning.
The substantial amount of prior works, \cite{blackhall2020calculation, deip2022, blackhall2022, liu2021grid, liu2022using, lankeshwara2022dynamic, bassi2022deliverables}, address the first two use case for DOE in Fig. \ref{fig:usecase}, i.e., concerning network awareness and maximizing DG integration of flexible resources while avoiding network congestion incidents.
\vspace{-5pt}

\subsection{\textcolor{black}{Related literature}}
Operating envelopes for power system flexibility are introduced in \cite{nosair2015flexibility}.
The research in the domain of DOE is widely used in Australian DNs \cite{blackhall2020calculation, deip2022} and gradually gaining traction worldwide where DG penetration in DNs is high.
In the USA, with the FERC order 2222, \cite{ferc}, allowing distributed energy resources (DERs) to participate in wholesale electricity markets, the relevance of DOEs is going to grow drastically, this is exhibited in the panel session at PES General Meeting 2022 \cite{link999}.
Recent literature showcases the applicability of DOEs in  building energy management \cite{gasser2021predictive}, and demand response \cite{lai2022demand}.
The fairness attribute of DOEs is dealt with in 
\cite{petrou2020operating, yi2022fair}.
Liu and Braslavsky in \cite{liu2022robust} use centralized optimal power flow (OPF) based robust DOE calculation.
A framework for calculating probabilistic operating envelopes is introduced in \cite{yi2022operating}.
The dynamic active (P) and reactive (Q) power DOE calculation using centralized OPF is elaborated in 
\cite{gerdroodbari2022dynamic}.

These DOEs are similar to P-Q charts, which essentially define the feasible operating region for flexibility resources. Novel frameworks for P-Q charts are introduced in 
\cite{ivas2020pq, capitanescu2018tso, riaz2021modelling}.

\vspace{-5pt}
Many smart meters already measure the voltage magnitude at the point of common coupling (PCC). Prior works such as \cite{euniversal3,euniversal52} show that voltage violations precede line loading limit violations. Therefore, voltage magnitude is a strong indicator of DN congestion, thus, voltage magnitude measurement \textcolor{black}{and forecast is used} for DOE generation in this work. 

\begin{table*}[!htbp]
 \normalsize
	\caption {Centralized vs presented decentralized DOE generation} 
	\label{tab:centralizedvsdecentr}
	\begin{center}
		\begin{tabular}{p{88mm}|p{88mm}}
			\hline
			 \textbf{Centralized DOE calculation} & \textbf{Proposed decentralized DOE} \\
			\hline 
             \hline
            \includegraphics[width=0.45\columnwidth]{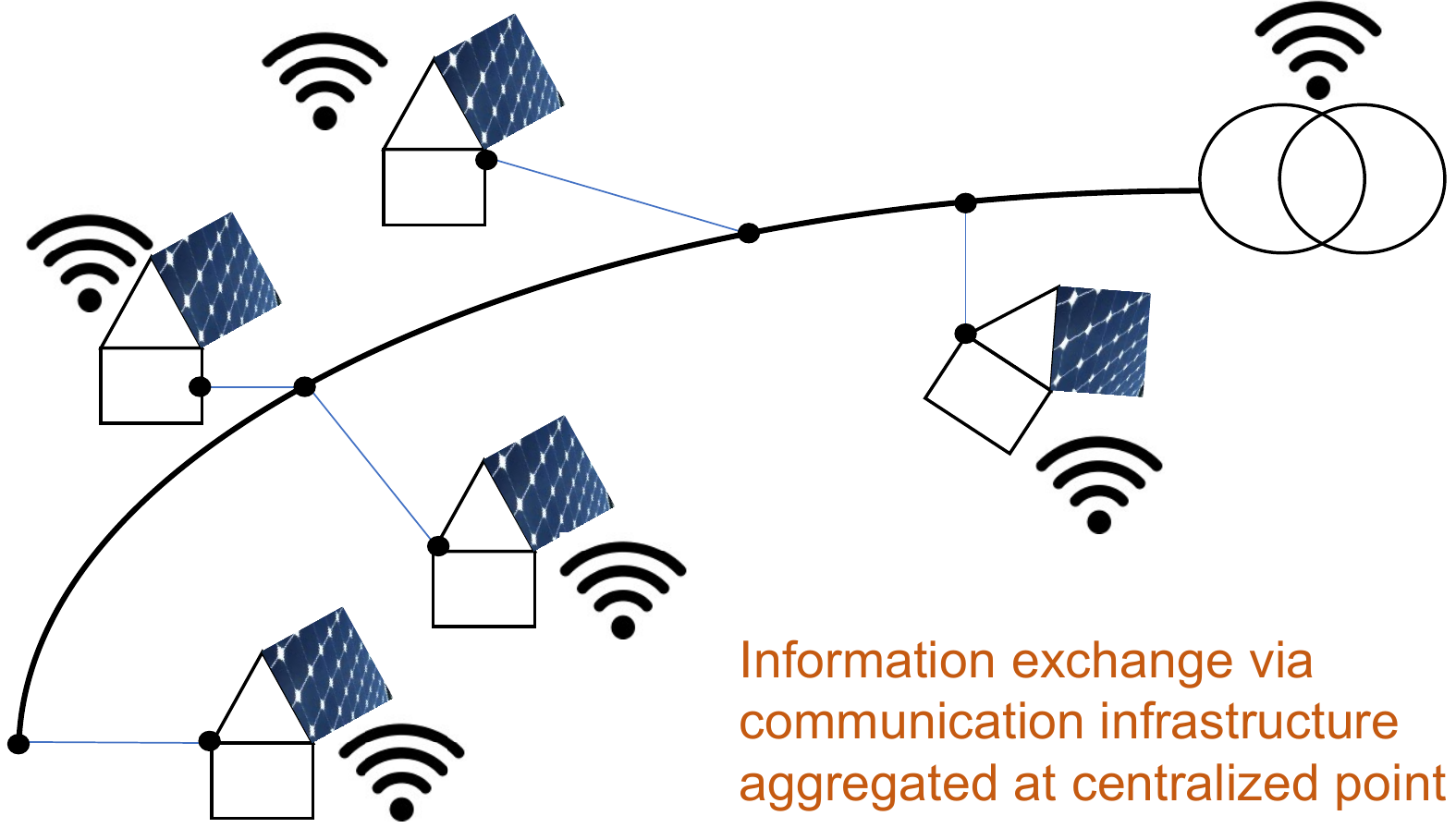}  & \includegraphics[width=0.45\columnwidth]{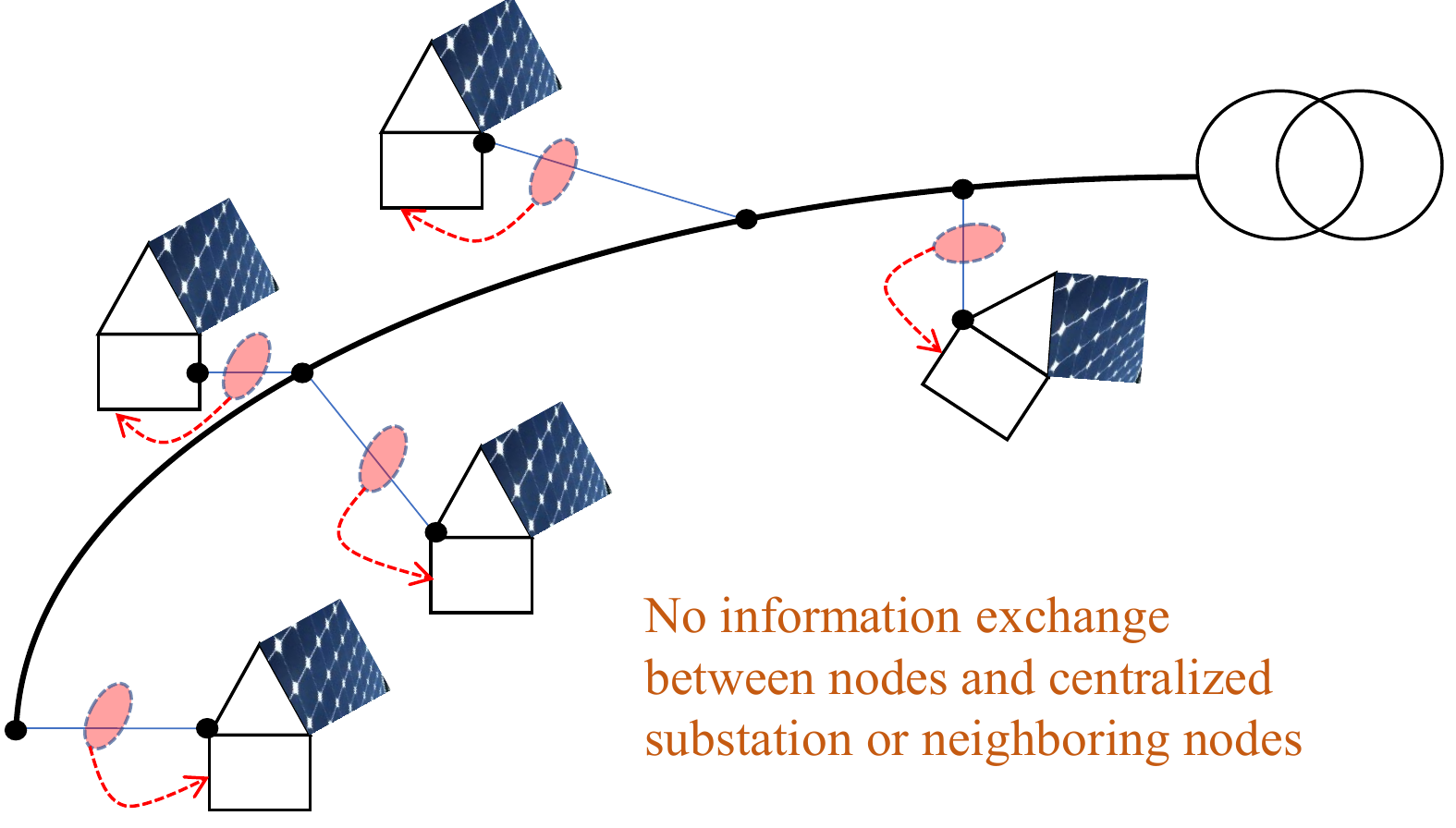}\\ 
             \hline
            $\bullet$ Communication infrastructure needed eg. 4G, PLC etc & $\bullet$ \textcolor{black}{No communication needed as only voltage magnitude at PCC is used}\\
            $\bullet$ Centralized (optimal) power flows are used for DOE calculation \cite{liu2022using, liu2022robust}  & $\bullet$ \textcolor{black}{Con: Lack of centralized coordination, may lead to convergence issues leading to over and/or under compensation of corrective action} \\
            $\bullet$ High observability is needed for measurement feedback $\forall$ the nodes  & $\bullet$ Suitable for low observability networks: distribution/transmission \\
            $\bullet$ Prone to communication delays, cyberattacks, missing information & $\bullet$ Cyber-secure \& resilient to communication failures, missing data \\
            $\bullet$ Higher accuracy for the calculation of DOE. Simultaneous coordination of resources can be performed in case of centralized intervention. & $\bullet$ Suitable for \textit{autonomous operation} for applications such as: (a) peer-to-peer (P2P) trading without centralized interference \& DN awareness, (b) decentralized flexibility activation for congestion mitigation \\
            $\bullet$ Accurate information of network topology is crucial, & $\bullet$ Network topology info (line impedance, phase connectivity): not needed. \\
            \hline
        \end{tabular}
		\hfill\
	\end{center}
\end{table*}

\subsection{\textcolor{black}{Contributions and organization of the paper}}
The goal of this paper is to provide a novel framework for calculating DOEs, using only \textcolor{black}{local voltage magnitude measurement and/or forecast at the PCC.}
A summary of the pros and cons of DOE calculation using centralized versus the \textcolor{black}{presented (partially)} decentralized DOE calculation is summarized in Tab. \ref{tab:centralizedvsdecentr}.
The key contributions of this paper are:\\ \textcolor{black}{
$\bullet$ \textit{Real-time DOEs using only voltage measurement}: 
To the best of our knowledge, this is the first work that proposes a decentralized DOE calculation without any centralized feedback.\\
$\bullet$ \textit{Time-ahead DOEs using voltage forecast}: we use centralized power flows for generating voltage forecast. Two methods for calculating DOEs are presented: M1 applies chance constraint on the voltage values and calculates the DOE  for each node and time, \textcolor{black}{and M2} applies chance constraint on the envelopes calculated based on each of the scenarios of voltage. Numerically, we observed that M1 is more than two times faster computationally compared to M2, while substantially reducing the data communication needs.\\
$\bullet$ \textit{Robust DOE calculation framework} \textcolor{black}{is proposed} using nodal voltage forecast in time-ahead setting. A chance constraint (CC) level is implemented analytically, similar to \cite{santos2020stochastic, tagawa2017weighted, petsagkourakis2022chance}. The CC avoids over-conservative envelopes that would be restrictive for local objectives of the flexibility devices, \textcolor{black}{such as (a) energy cost, (b) loss cost minimization, (c) energy market revenue, (d) social welfare, (e) quality-of-service maximization, etc.}\\
$\bullet$ Numerically, we show that day-ahead DOE calculated temporally resembles the flexibility needs assessment (FNA) identified in \cite{hashmi2022chance}, implying the efficacy of \textcolor{black}{the presented} framework. Furthermore, we extend the calculated DOEs to form nodal and temporal P-Q charts that also consider (a) power factor limit \cite{hashmi2020arbitrage}, and (b) the capacity of the inverter.
}

This paper is organized as follows.
Section \ref{section2}, describes the formation of robust dynamic operating envelopes for network awareness.
Section \ref{section3}, details the numerical case studies.
Section \ref{section4} concludes the paper.

\pagebreak

\section{Methodology for network awareness}  
\label{section2}
\begin{figure}[!htbp]
	\centering
	\includegraphics[width=6.1in]{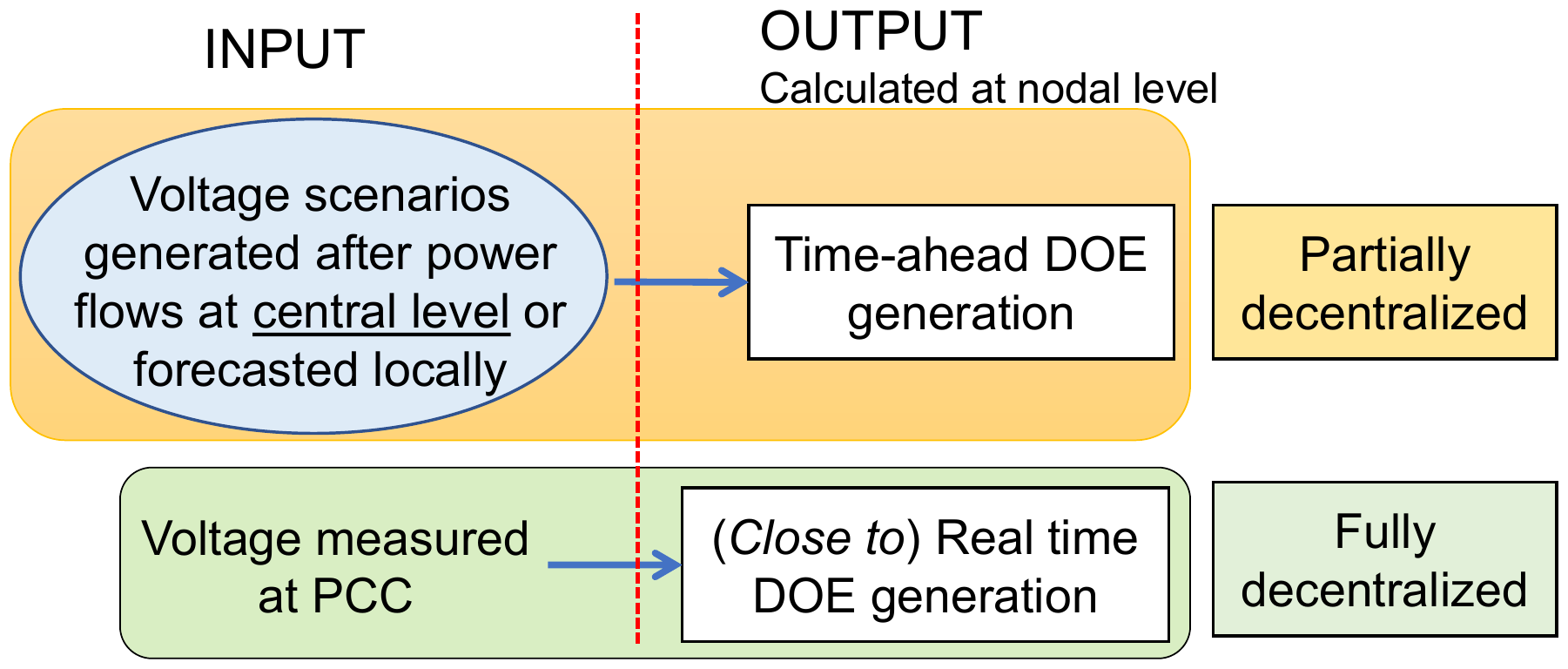}
	\caption{{Time-ahead and real-time DOE generation framework} }
	\label{fig:blockdiagram}
\end{figure}

In this work, a decentralized \textcolor{black}{and a partially decentralized} mechanism for generating DOEs \textcolor{black}{is proposed} that takes as input the voltage measurement in real-time
\textcolor{black}{and for time-ahead setting, power flows are performed for different load scenarios depending on the forecast of load and distributed generation and associated forecast errors, refer to \cite{hashmi2022chance} for details on scenario generation. 
We present two models for generating DOEs in a time-ahead setting, requiring different levels of communicated data.  Note that in
time ahead time setting, communication latency is not a concern
as small delays will not impact DOE generation.
Fig. \ref{fig:blockdiagram} shows the inputs and output of the DOE generation proposed.
}

%
The proposed DOE builds on our prior work on projecting traditional \textcolor{black}{volt-var} and volt-watt inverter policies for identifying the feasible range for operating resources \cite{9956913}. In \cite{9956913} we identify an active and reactive power range selection policy that reduces the locational discrepancy of such flexible resources.
This work proposes a methodology for generating DOE for flexibility services, considering local DN constraints. These DOE subsequently be used for optimizing operational states of flexible resources based on the service performed in the energy market(s). 
The proposed \textcolor{black}{real-time or close to real-time DOE (RT-DOE) generation framework} for a node is a function of nodal voltage levels at the PCC. In many DNs the network topology is not accurately known, for such DNs, the proposed RT-DOE generation mechanism can be directly applied as only PCC measurement of voltage is required. Further, since RT-DOE is generated locally, no additional communication is needed, making the proposed DOE generation inherently resilient to cyberattacks. 
\textcolor{black}{However, for time-ahead DOE generation nodal voltage forecast is needed. In this work, power flows are performed at the centralized level for different loading scenarios and the voltage forecast is communicated to each node. DOEs are generated locally based on these voltage levels communicated.
Although voltage forecast scenarios in this work are calculated centrally, local and distributed voltage forecasting methods can also be utilized.}


Next, we describe the DOE generation framework. Subsequently, two methods of generating robust envelopes \textcolor{black}{is proposed} while considering a CC or risk level. The CC is used in order to avoid too conservative solutions, which would provide little to no room for optimizing flexible resources.

\begin{figure}[!htbp]
	\centering
	\includegraphics[width=5.6in]{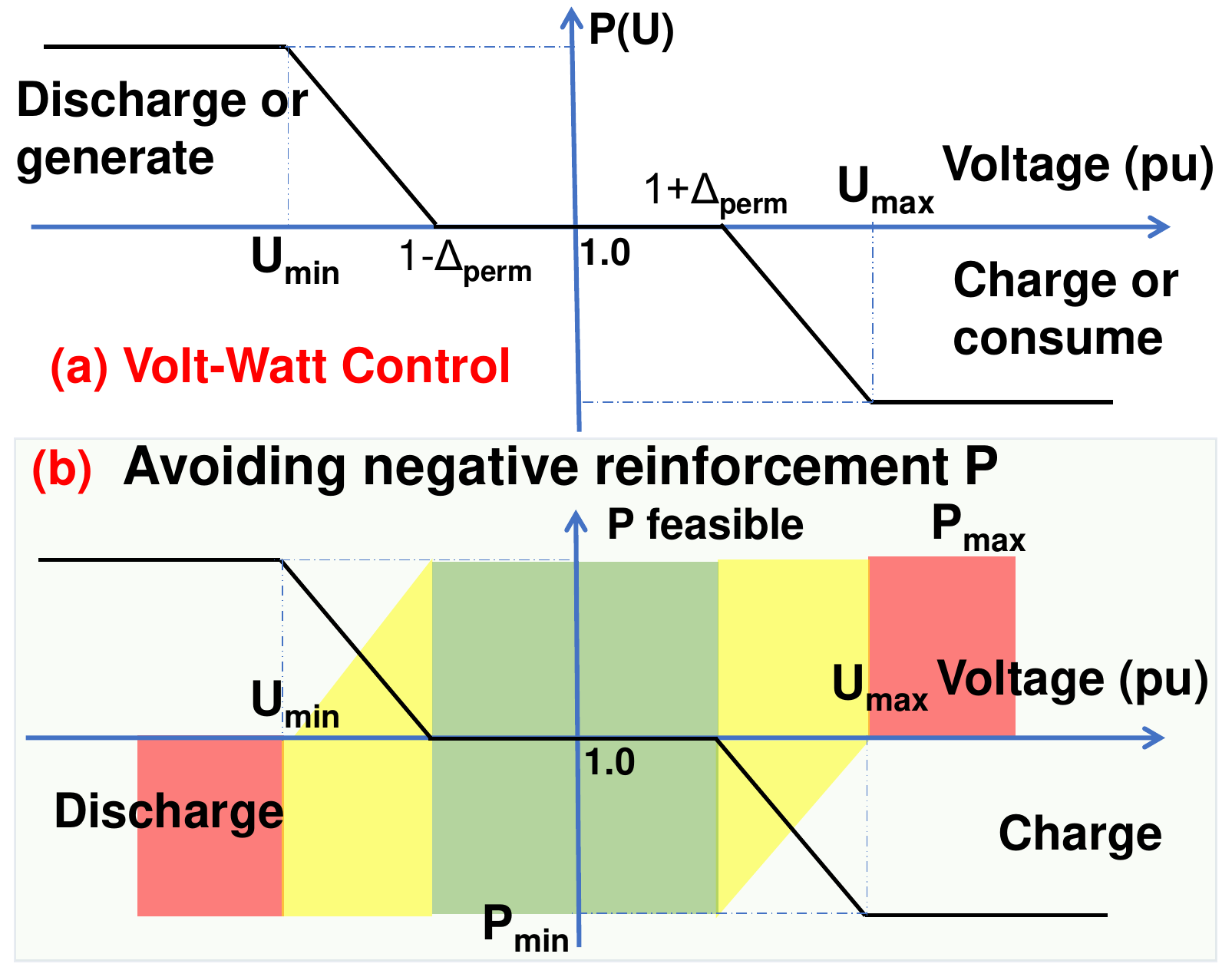}
	\caption{{Active power DOE based on \cite{9956913}. (a) Volt-watt or P(U) control and (b) envelope for \textcolor{black}{active power,} P. $U_{\min}$   and $U_{\max}$  denotes minimum and maximum voltage levels allowed,  $\Delta_{\text{perm}}$ denotes permissible voltage fluctuation, $P_{\min}, P_{\max}$ are the minimum and maximum active power output and 
$Q_{\min}, Q_{\max}$ are the minimum and maximum reactive power output, 
$pf$ denotes the power factor lower limit.} }
	\label{fig:penvelope}
\end{figure}

\begin{figure}[!htbp]
	\centering
	\includegraphics[width=5.6in]{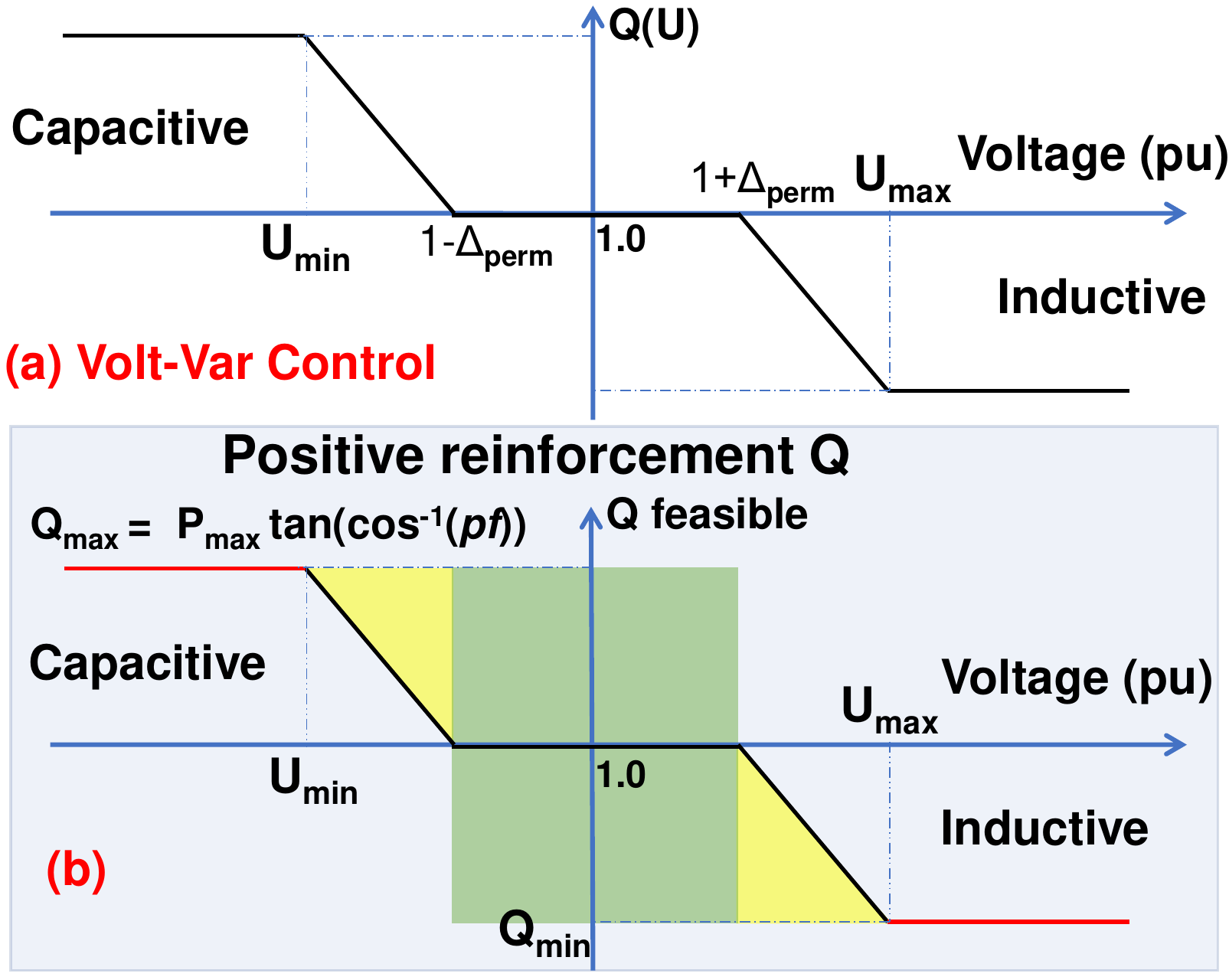}
	\caption{{Reactive power DOE. (a) \textcolor{black}{volt-var} or Q(U) control and (b) envelope for \textcolor{black}{reactive power,} Q \cite{9956913}.} }
	\label{fig:qenvelope}
\end{figure}

\subsection{Dynamic operating envelopes}
\label{sectionAA}
\textcolor{black}{P(U) and Q(U) inverter control are also referred to as volt-watt and volt-var control.}
The use of P(U) and Q(U) inverter control in standalone and/or in combination are popular practices for the operation of radial DNs. It uses the drooping behavior to limit excess active and reactive power injection (capacitive) or consumption (inductive) thus, not aggravating grid voltage further. 
In \cite{9956913}, two adaptations of \textcolor{black}{volt-var} and volt-watt inverter policies \textcolor{black}{are proposed}, referred to as \textit{positive reinforcement control} (PRC) and \textit{avoiding negative reinforcement control} (ANRC).
PRC outputs more restrictive operating envelopes compared to  ANRC.
We observed that ANRC projections used for active power (P) and PRC used for setting reactive power (Q) set points reduce the locational discrepancy \cite{9956913}. In this work, we will extend the hybrid projections used for P and Q for generating DOE.
Fig. \ref{fig:penvelope} and Fig. \ref{fig:qenvelope} show the range selection mechanism used for generating DOE.
The feasible region is shown as traffic-light control with green as no reduction in the feasible region, yellow shrink the feasible region, and red shrink it even further, see Fig. \ref{fig:penvelope}(b) and Fig. \ref{fig:qenvelope}(b).
The volt-watt and \textcolor{black}{volt-var} plots shown in Fig. \ref{fig:penvelope}(a) and Fig. \ref{fig:qenvelope}(a). 
The operating region for the inverter is defined based on the voltage magnitude $U_{i,t}$. For active power, we denote the region as $R_P(U_{i,t})\equiv [ R^{U_{i,t}}_{P_{\min}}, R^{U_{i,t}}_{P_{\max}}]$, and for reactive power, as $R_Q(U_{i,t}) \equiv [R^{U_{i,t}}_{Q_{\min}}, R^{U_{i,t}}_{Q_{\max}}]$.
where $ R^{U_{i,t}}_{P_{\min}}$, $ R^{U_{i,t}}_{Q_{\min}}$ denote the lower operating envelopes and $ R^{U_{i,t}}_{P_{\max}}$ and $ R^{U_{i,t}}_{Q_{\max}}$ denote the upper operating envelopes for P and Q respectively. Note that these ranges are modified based on the zone, the voltage magnitude and control type.
\textcolor{black}{Fig. \ref{fig:penvelope}(b) and Fig. \ref{fig:qenvelope}(b) are directly used for P and Q DOE generation for real-time setting based on the voltage measurement at PCC.}

\subsection{Generating of robust \textcolor{black}{time-ahead DOEs}}
\label{sec_subb}
In Sec. \ref{sectionAA} we detail the DOE generation using 
voltage measurement. This method does not consider voltage fluctuations due to forecast errors. In the time ahead flexibility planning, parameter uncertainty needs to be considered \textcolor{black}{in the form of load and DER scenarios for centralized power flow. 
The voltage magnitude time series for these scenarios is then communicated to the nodes (partially in model M1 and fully in M2).
For M1, an empirical cumulative distribution function (ECDF) is identified for nodal voltage distribution for each time instant.}
We use non-parametric ECDF for risk assessment \cite{hashmi2022chance}.
Envelopes generated for the absolute worst case of probable voltage fluctuation would indeed make the flexible resource operation robust. However, such realizations of the worst case are highly improbable. This motivates us to use CC. 
Based on the distribution of data, different levels of robust solutions can be created.  The CC for this robust formulation is analogous to the risk. It \textcolor{black}{is} governed by the incentive of participating and the penalty for not ensuring healthy network operation.
In this work, we assume the CC level is fixed at 5\%. 
\begin{figure}[!htbp]
	\centering
	\includegraphics[width=0.92\columnwidth]{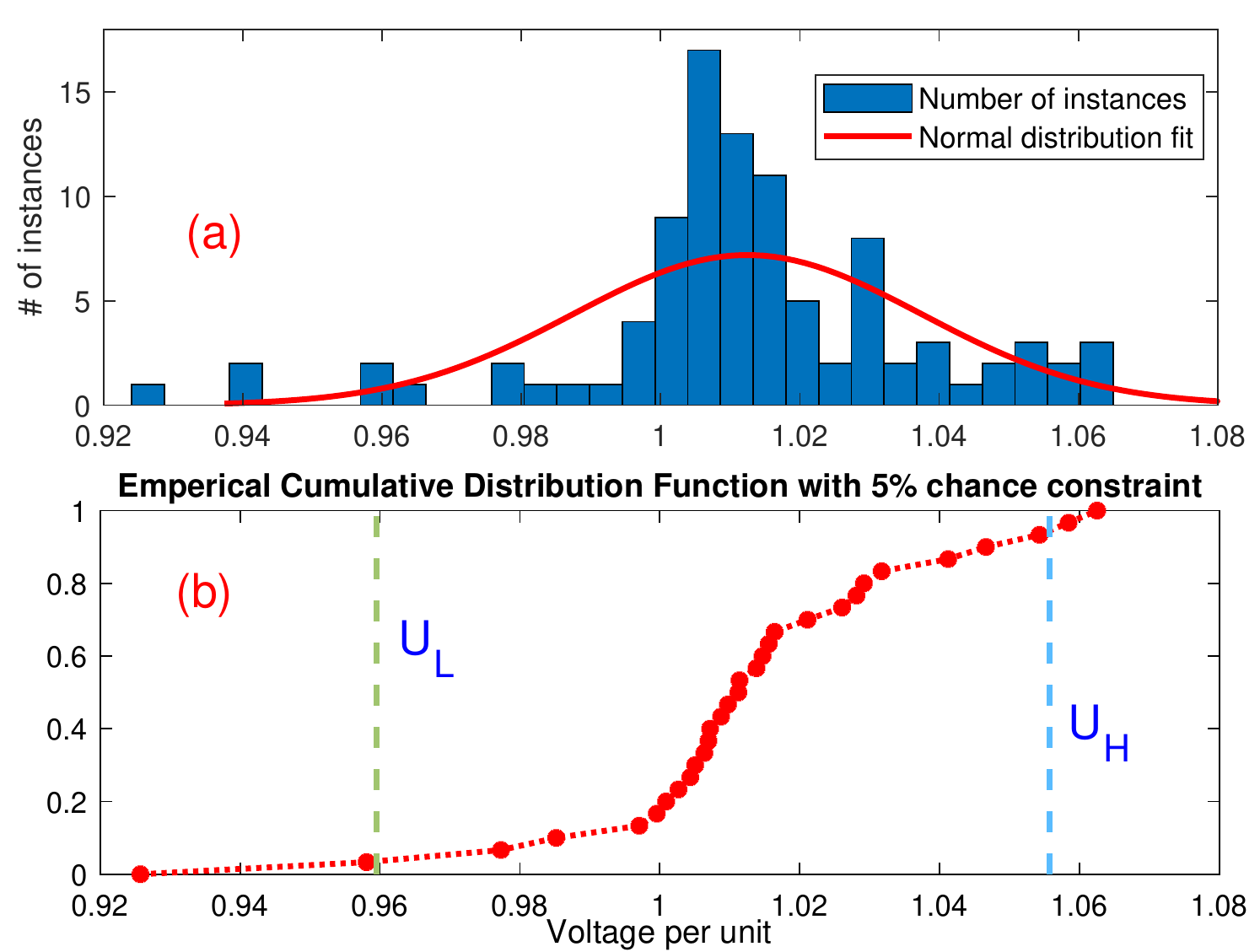}
	\caption{{Nodal voltage statistical attributes. (a) the histogram and an approximately normal distribution fit for nodal voltage scenarios, and (b) ECDF for all the nodal voltage scenarios.} }
	\label{fig:nonparametr}
\end{figure}

We propose two methods, M1, M2 for calculating risk-averse robust DOEs based on the sequence of CC implementation.

\subsubsection{M1: Apply CC on voltage levels}
\begin{figure}[!htbp]
	\centering
	\includegraphics[width=6in]{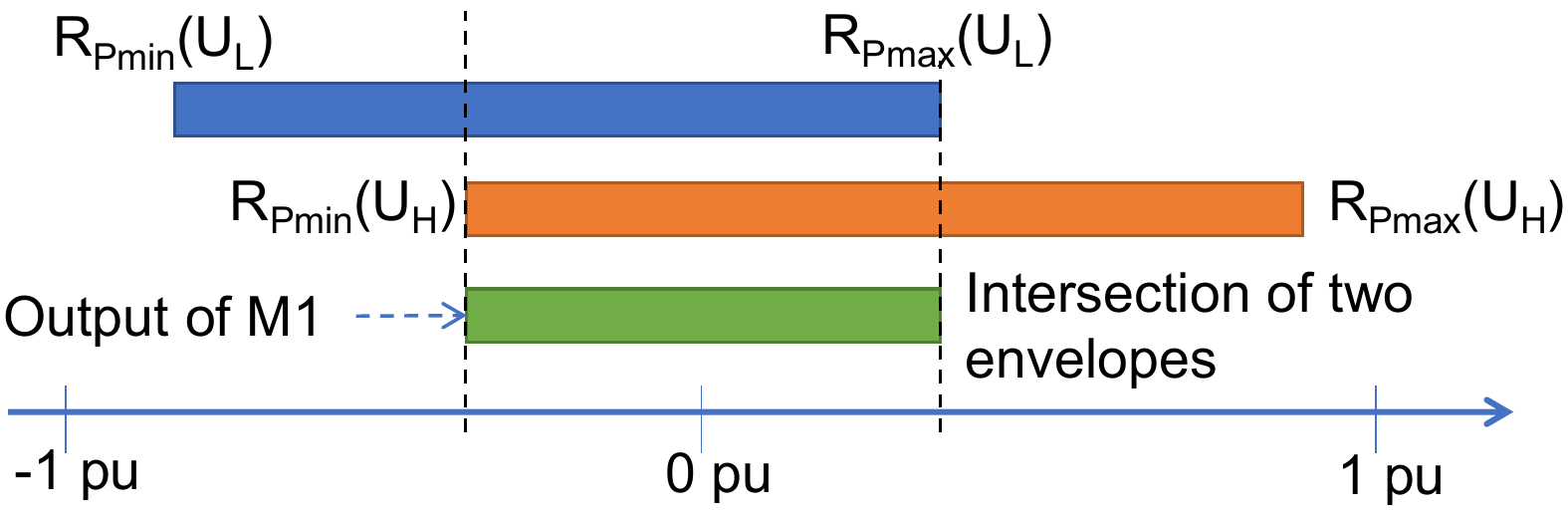}
	\caption{ {The upper and lower levels of voltage are used for envelope generation. The intersection of these envelopes is the output for the M1 method of robust envelope generation.} }
	\label{fig:m1flow}
\end{figure}

In this method, the CC level is applied to all the voltage scenarios for each node and time.
\textcolor{black}{Based on the voltage distribution shown in Fig. \ref{fig:nonparametr}(a), ECDF is identified as shown in Fig. \ref{fig:nonparametr}(b). $U_L$ and $U_H$ voltage time series are communicated to each node for the look-ahead time horizon.
These risk-averse voltage levels are used to calculate DOEs.
}
Fig. \ref{fig:m1flow} summarizes the M1 method\textcolor{black}{, where time-ahead DOE is the intersection of DOEs for inputs $U_L$ and $U_H$}.

\subsubsection{M2: Apply chance constraint on DOE}
\begin{figure}[!htbp]
	\centering
	\includegraphics[width=6in]{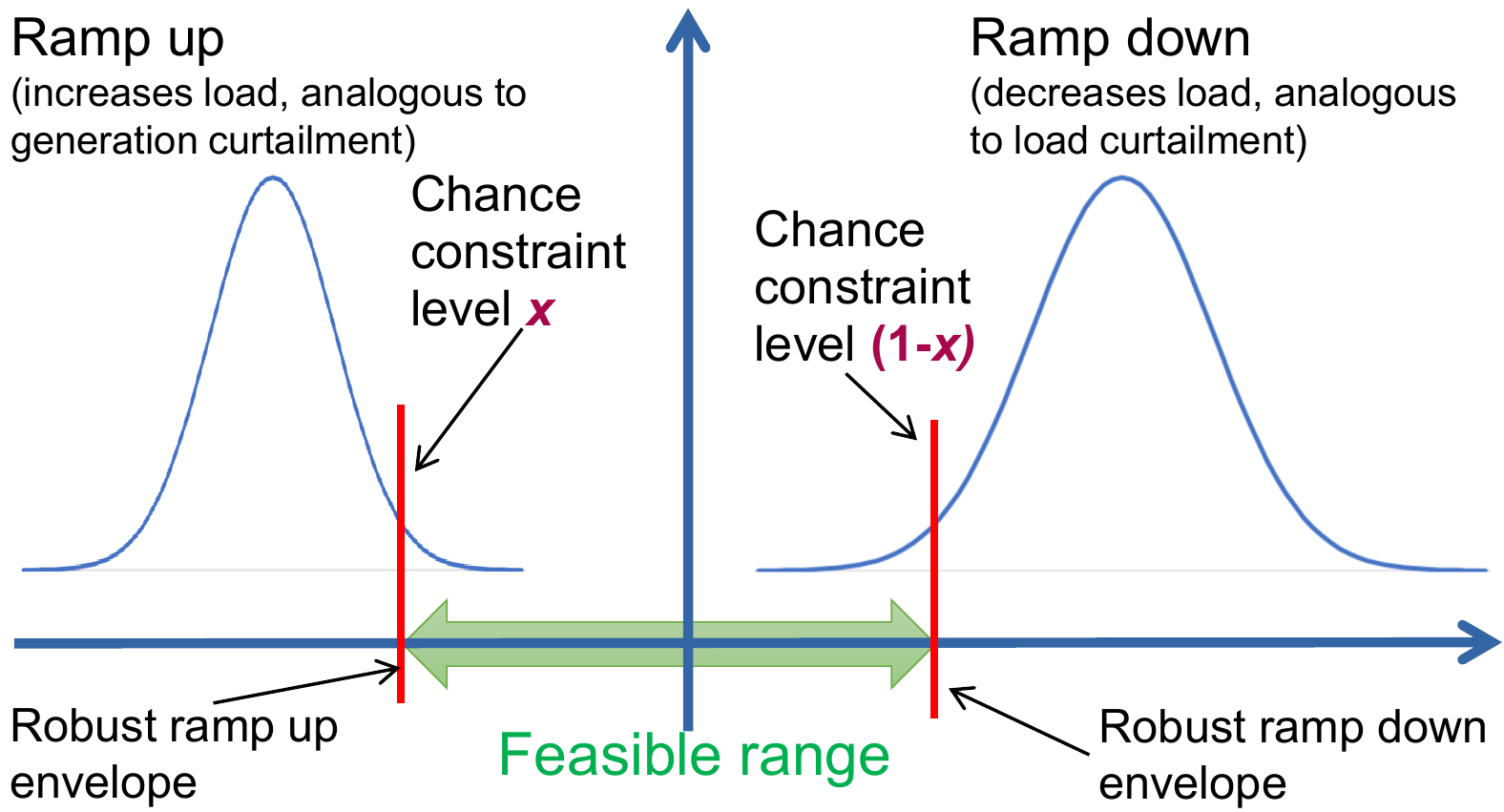}
	\caption{ {Feasible range calculated using M2} }
	\label{fig:flexneed}
\end{figure}

In this method, DOE is calculated for all scenarios for each time and node. 
\textcolor{black}{Thus, all voltage time series for all scenarios needs to be communicated to the prosumer node.}
The ramp up and ramp down P and Q robust DOE is identified by applying CC level, as shown in
Fig. \ref{fig:flexneed} for the DOE calculated.

\textcolor{black}{For centralized voltage scenario calculation, M1 compared to M2 DOE generation would be more efficient as only $U_L$ and $U_H$ time-series needs to be communicated to the prosumer node.}

\subsection{Performance indices}
Numerical results use the following performance indices
\begin{itemize}
    \item For comparing M1 and M2:
    \begin{enumerate}
        \item \underline{Sum of absolute error} in DOEs generated by methods M1 and M2, given as
    \end{enumerate}
    \end{itemize}
    \begin{gather}
    \label{eq:eq1}
            \Delta P_E \text{=}\sum_t \sum_i \{ | R_{P_{\min}}^{\text{M1}, i,t} - R_{P_{\min}}^{\text{M2}, i,t} | + | R_{P_{\max}}^{\text{M1}, i,t}- R_{P_{\max}}^{\text{M2}, i,t} | \},\\
            \Delta Q_E \text{=}\sum_t \sum_i \{ | R_{Q_{\min}}^{\text{M1}, i,t} - R_{Q_{\min}}^{\text{M2}, i,t} | + | R_{Q_{\max}}^{\text{M1}, i,t}- R_{Q_{\max}}^{\text{M2}, i,t} | \}.
            \label{eq:eq2}
        \end{gather}
        \begin{enumerate}\addtocounter{enumi}{1}
            \item \underline{Computation time} in seconds.
        \end{enumerate}
        \begin{itemize}
    \item \underline{Comparing DOE and FNA}: For comparing the decentralized DOE framework proposed compared to the centralized method, we use the flexibility needs assessment output generated based on \cite{hashmi2022chance}. It is assumed that the shrinking of the feasible region of DOE is an indicator of the flexibility needed by DN.
\end{itemize}


\textcolor{black}{
\subsection{Algorithm for DOE calculation}
Algorithm \ref{alg:nodalsensitivity} outlines the procedure to identify DOEs in time-ahead and close to real-time frames. 
}
\begin{algorithm}
	{\textbf{Inputs}}: Nodal voltage forecast, voltage measurement \\
    {\textbf{Output}}: Dynamic operating envelopes in day-ahead \& real-time, 
	\begin{algorithmic}[1]
	    \State Generate day-ahead dynamic operating envelopes (DA-DOE),
     \begin{itemize}
         \item Perform power flows \& communicate nodal voltage scenarios
         \item Use these voltage values for generating the DA-DOE based on M1 and M2 models detailed in Section \ref{sec_subb}.
     \end{itemize}
        \State Calculate DOEs in real-time (RT-DOE) using voltage measurement at PCC. DOE calculation is based on Section \ref{sectionAA}.
	\end{algorithmic}
	\caption{\texttt{DOE in time-ahead \& real-time}}
	\label{alg:nodalsensitivity}
\end{algorithm}

\pagebreak

\section{Numerical results}
\label{section3}
\begin{figure}[!htbp]
	\centering
	\includegraphics[width=0.79\textwidth]{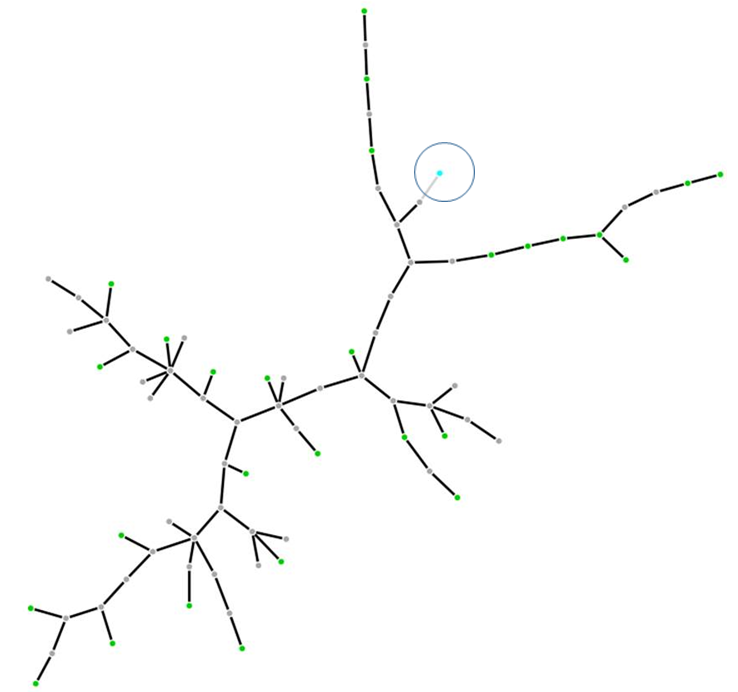}
	\caption{Network diagram \cite{hashmi2023estimate} for the 76 bus Spanish DN. The substation is indicated with a blue circle.}
	\label{fig:networkdiag}
\end{figure}

The DN considered is an adaptation of one of the Spanish LV feeders described in \cite{koirala2020non}.
The DN consists of 76 nodes and 75 branches connecting the nodes.
{52 loads are connected in 28 DN nodes.}
The DN diagram is shown in Fig.~\ref{fig:networkdiag}.
\textcolor{black}{The DN used in this work along with load profiles can be found on GitHub \cite{githubfna}.}


\subsection{Case study 1: Methods for DOE calculation}
{The case study outline is as follows:
\begin{itemize}
\item \textit{Objective}: Evaluation of DOE calculation methods M1, M2,
\item \textit{Data used}: 76 bus Spanish DN,
\item \textit{Methodology used}: Proposed in Section \ref{section2},
\item \textit{Outcome}: M1 is computationally better compared to M2.
\end{itemize}
}
\begin{figure}[!htbp]
	\centering
	\includegraphics[width=0.8\columnwidth]{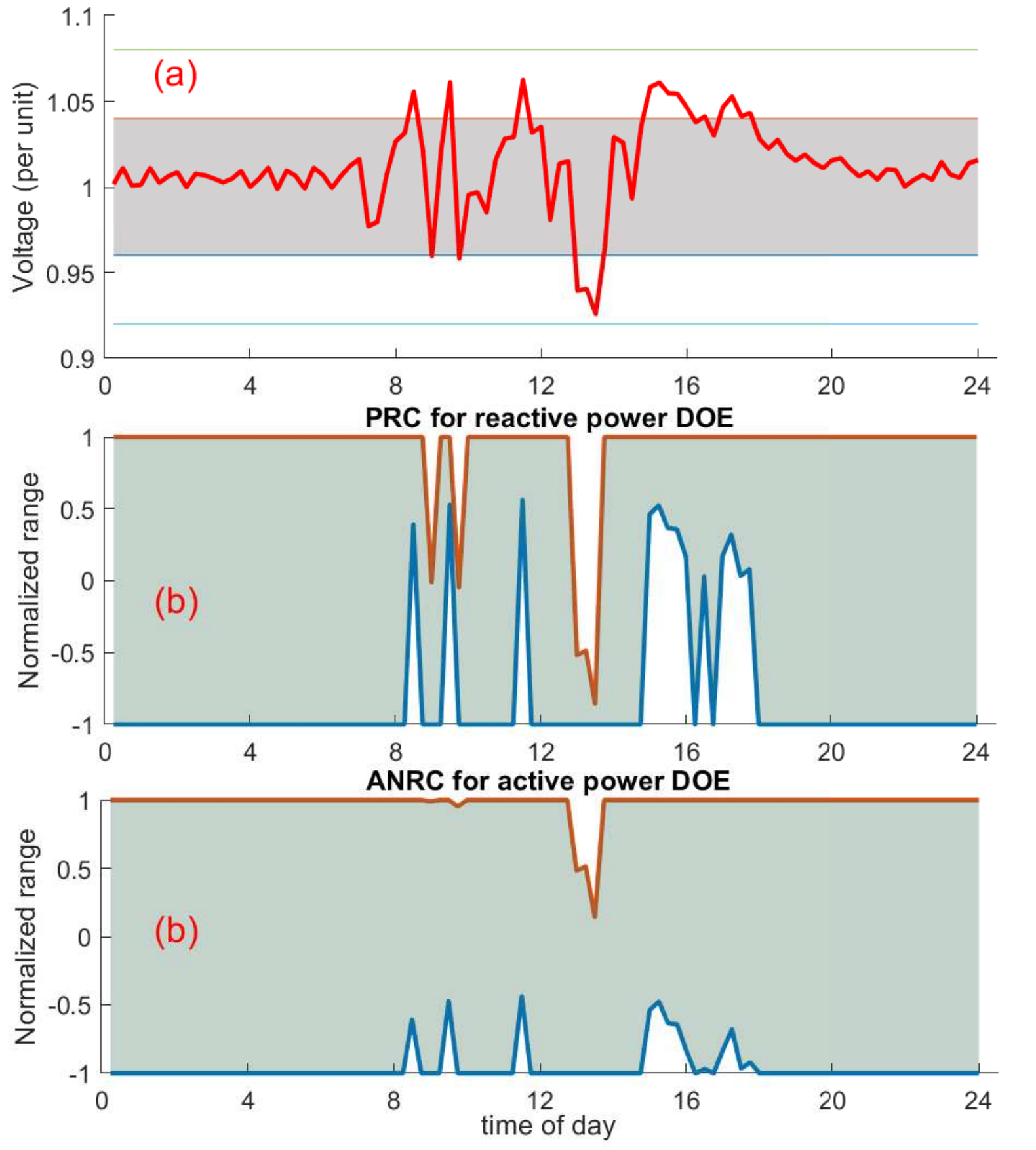}
	\caption{ {DOE for active and reactive power of a given node in real-time based on nodal voltage measurement time-series, shown in (a).
 (b) shows the reactive power DOE is based on PRC, and (c) shows the active power DOE based on ANRC in per unit value of flexibility.} }
	\label{fig:case11}
\end{figure}


The visualization for {P and Q DOE based on voltage time series measurement in \textit{real-time} \textcolor{black}{(RT-DOE)} is shown in Fig. \ref{fig:case11}}. 
The parameters used are $U_{\max} = 1.08, U_{\min} = 0.92, \Delta_{\text{perm}} = 0.04$, and ANRC is used for P DOE and PRC is used for Q DOE.
The shaded region in Fig. \ref{fig:case11}(b) and (c) shows the feasible region of operation.

\begin{figure}[!htbp]
	\centering
	\includegraphics[width=0.9\columnwidth]{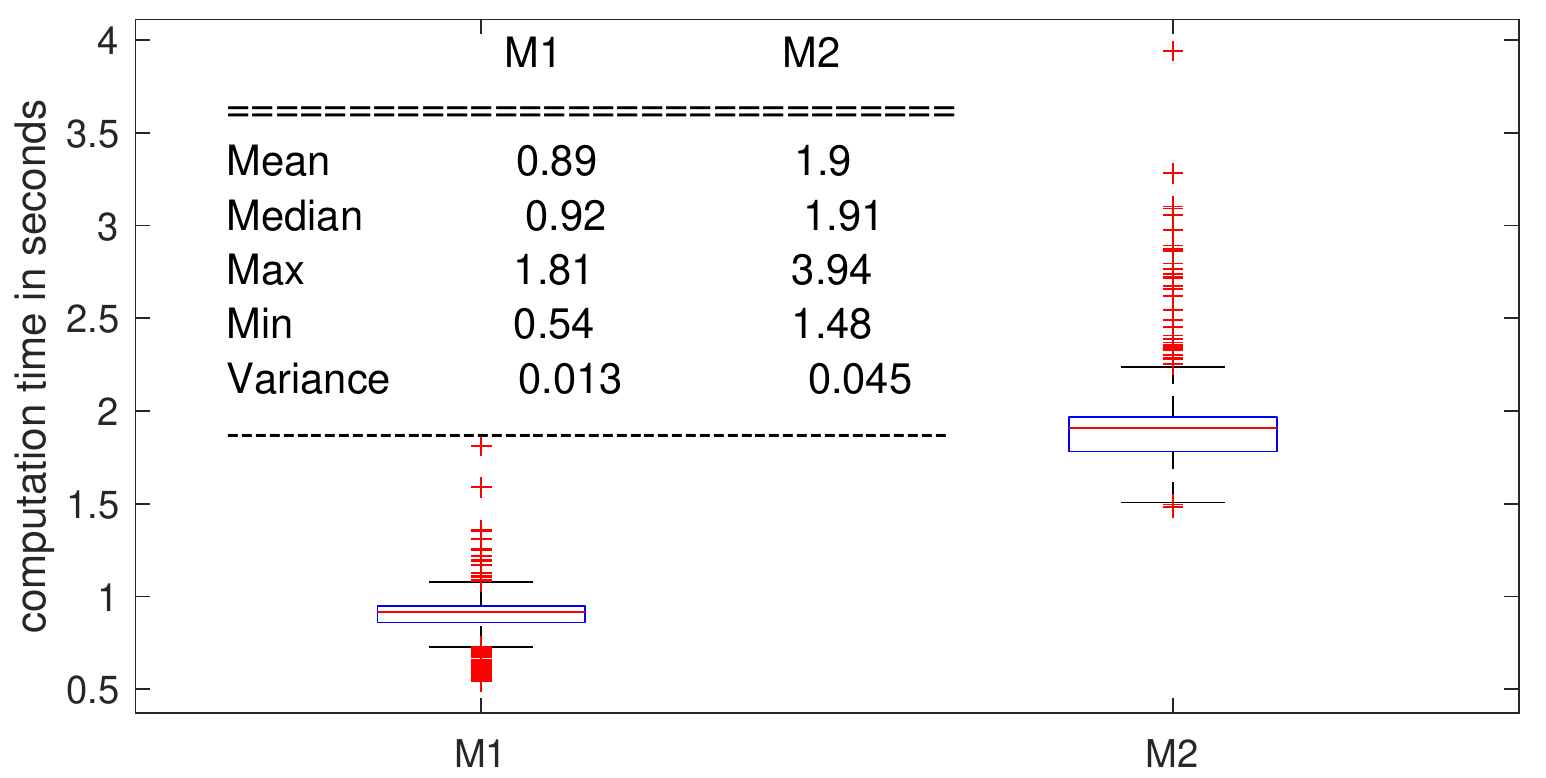}
	\vspace{-4pt}
	\caption{ {Comparison of computation time for models M1 and M2 for 1000 Monte Carlo simulations.} }
	\label{fig:case12}
\end{figure}

\textcolor{black}{DA-DOE's are calculated using voltage forecasted scenarios calculated at a centralized level. The voltage forecast scenarios are generated based on load forecast scenarios used for performing power flows. 1000 scenarios are generated for load profiles, see \cite{hashmi2022chance} for details on load profile scenarios.}
For comparing DA-DOE generating methods M1 and M2, we calculate the sum of absolute errors using \eqref{eq:eq1} and \eqref{eq:eq2}:
$\Delta P_E = 0$ and $\Delta Q_E = 0.23$.
The computation time comparison for 1000 Monte Carlo simulations \textcolor{black}{is} shown in 
Fig. \ref{fig:case12}.
Using, these performance indices, we can draw a conclusion that \underline{M1 is computationally at least 2 times better compared to M2}, while no significant difference is observed in the sum of absolute errors. 
\textcolor{black}{Further, M1 requires only 2 ($U_L$ and $U_H$) time series to be communicated to each node. On the contrary, M2 requires 1000 (for each scenario) times series to be communicated for each node.}
Thus, subsequently, we use model M1.

\vspace{-10pt}

\subsection{Case study 2: Comparison of DA-DOE with FNA}
\begin{figure}[!htbp]
	\centering
	\includegraphics[width=0.8\columnwidth]{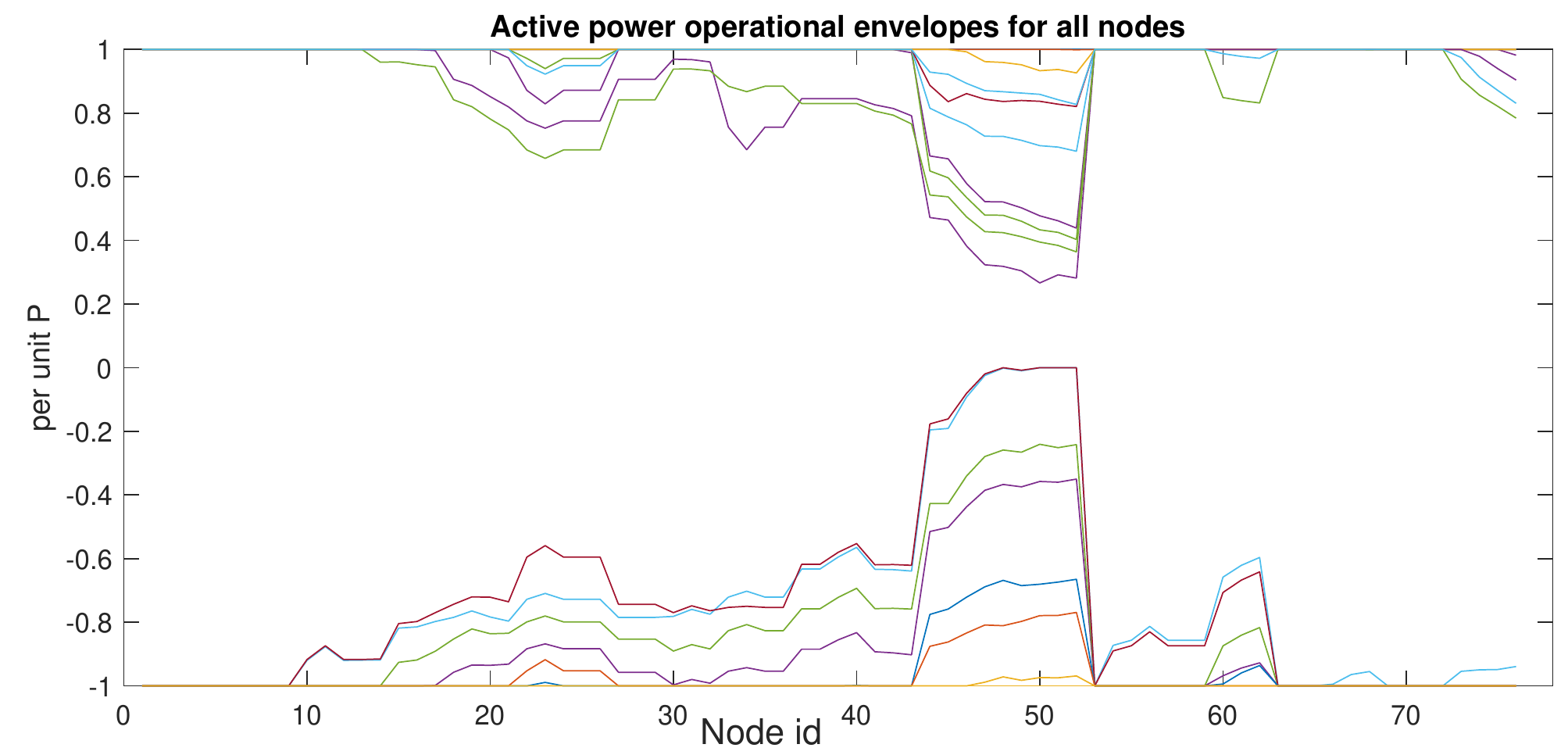}
	\vspace{-5pt}
	\caption{ {Nodal variation of the active power DOE for a 24-hour period.} }
	\label{fig:case21}
\end{figure}

\textcolor{black}{
The objective of this case study is to compare the DA-DOEs with day-ahead FNA identified in \cite{hashmi2022chance}.}
Clearly, from Fig. \ref{fig:case21}, the active power feasible ranges are more restrictive for some nodes compared to others, see nodes 48 to 52.
The reason for this inherent unfairness is due to the different voltage levels observed by the DN nodes.

\begin{figure}[!htbp]
	\centering
	\includegraphics[width=0.7\columnwidth]{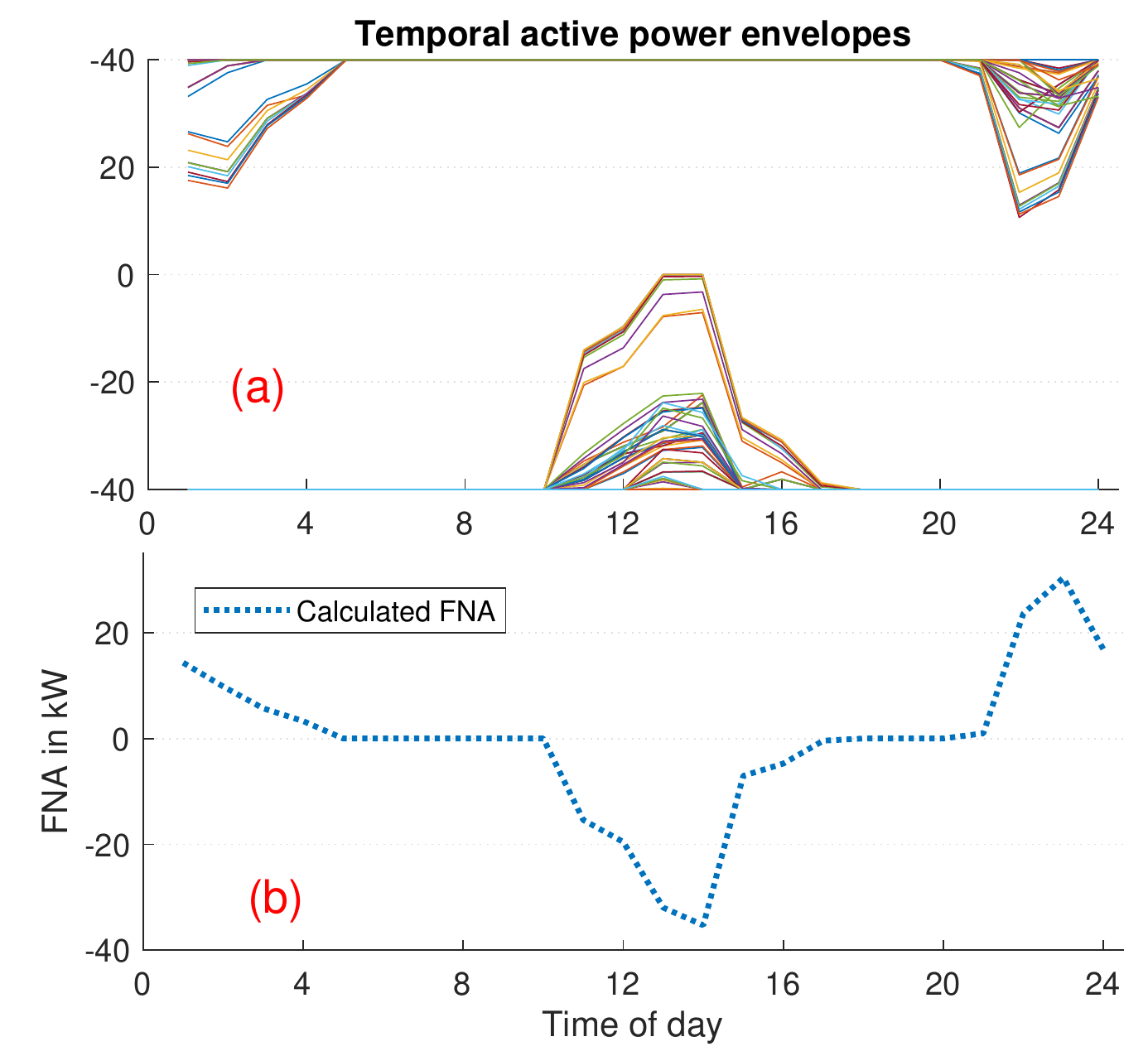}
	\vspace{-5pt}
	\caption{ {Comparison of DA-DOE with FNA calculated based on FNA-OPF proposed in \cite{hashmi2022chance}. (a) shows the temporal nodal DA-DOE, (b) shows the temporal FNA for the 76-bus DN.} }
	\label{fig:case22}
\end{figure}

Note that DA-DOEs are a preventive strategy and FNA is a corrective framework, pointing towards corrective action.
Fig. \ref{fig:case22}(a) the temporal DOEs and Fig. \ref{fig:case22}(b) shows the temporally aggregated nodal FNA. Clearly, the time periods where flexibility is needed are coinciding with the shrinking of DOEs implying the \underline{need for preventive action} from such resources.

\subsection{Case study 3: PQ charts}
\begin{figure}[!htbp]
	\centering
	\includegraphics[width=0.86\columnwidth]{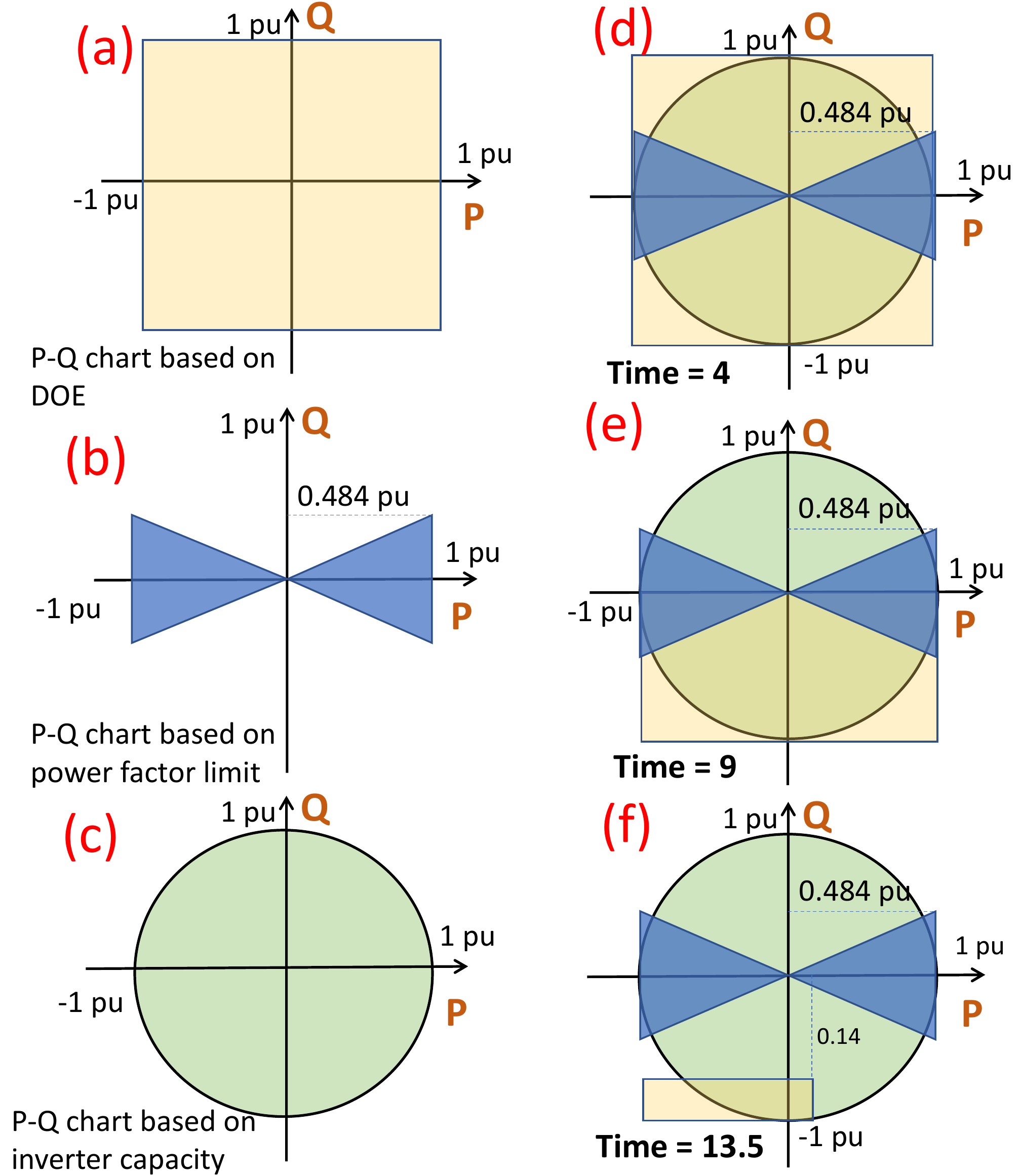}
	\vspace{-3pt}
	\caption{ {P-Q chart considering DOE and power factor limit and inverter size limit. (a) P-Q plot feasible space based on (a) DOE, (b) power factor limit of 0.9, and (c) inverter size of 1 per unit. Plots (d) to (f) denotes the P-Q charts for 3 different time instants on Fig. \ref{fig:case11}.} }
	\label{fig:pqchart}
\end{figure}

Fig. \ref{fig:pqchart} shows feasible spaces in shaded regions. In (a) the P-Q chart based on calculated DOE is shown.
(b) shows the feasible space due to power factor constraint, which limits the ratio of reactive and active power. For this demonstrative example, the power factor limit of 0.9 is considered.
(c) shows the feasible space due to the converter limit.
Plots Fig. \ref{fig:pqchart}(d), (e) and (f) show the P-Q charts for three-time instants of Fig. \ref{fig:case11}.
Note that for plot Fig. \ref{fig:pqchart}(f) the power factor limit could lead to a null set with no feasible region.


The P-Q charts thus identified can be used for LV flexible resources to contribute towards solving upstream network issues. These charts can be used for \underline{TSO-DSO coordination for flexible resource sharing}.

\pagebreak

\section{Conclusions and future directions}
\label{section4}
\textcolor{black}{Risk-averse robust partially decentralized and  fully decentralized framework in the context of time-ahead and real-time frames, respectively, for calculating dynamic operating envelopes (DOEs) is presented.
Real-time DOEs (RT-DOE) are calculated based on local voltage measurement, with no additional information exchange between neighboring nodes or centralized substations, thus eliminating the need for communication infrastructure.
Due to the decentralized RT-DOE calculations, the framework is inherently resilient to cyberattacks, communication failures, missing data, low levels of DN observability, and lack of network topology information.
For day-ahead DOE (DA-DOE) calculation framework, the voltage signals are communicated to the nodes, where DOEs are calculated, thus partially decentralized. 
}
In numerical evaluations, we apply the proposed DA-DOE calculation frameworks. Using performance indices, we show that applying risk-averse levels in the form of analytical implementation of chance constrained on voltage scenarios followed by DOE calculation leads to similar results when compared with chance constraints applied on all DOEs for all voltage scenarios.  
We also numerically extend the DOEs calculated to form P-Q charts considering power factor and converter capacity limits.

\textcolor{black}{
In future works, 
we will quantify the optimality gap between decentralized DOEs with centralized DOEs in the real-time frame. Subsequently, we will apply
the DOE calculation frameworks for the autonomous operation of distribution networks and industrial flexibility for solving congestion incidents and energy market participation. Furthermore, we will also apply DOEs for decentralized network-aware peer-to-peer trading.}

\pagebreak

\bibliographystyle{IEEEtran}
\bibliography{reference}

\end{document}